\documentstyle [a4p,epsf,12pt] {article}
\catcode`@=11 
\def \gsim{\mathrel{\mathpalette\@versim>}}
\def \lsim{\mathrel{\mathpalette\@versim<}}
\def \@versim#1#2{\lower0.4ex\vbox{\baselineskip\z@skip\lineskip\z@skip
     \lineskiplimit\z@\ialign{$\m@th#1\hfil##\hfil$%
     \crcr#2\crcr\sim\crcr}}}
\catcode`@=12 
\def\Zzero{\mathrm{{{\rm Z}^0}}}

\def\ra{\rightarrow}
\def\arr{\rightarrow}
\def\BPLUS{\mathrm{B^+}}

\def\Bq{\mathrm{B_q^0}}
\def\Bqb{\mathrm{\bar B_q^0}}
\def\Bd{\mathrm{B_d^0}}

\def\Bs{\mathrm{B_s^0}}

\def\BdBd{\mathrm{B_d^0 - \bar B_d^0}}
\def\BsBs{\mathrm{B_s^0 - \bar B_s^0}}

\def\DSPM{\mathrm{D^{*\pm}}}
\def\ccbar{\mathrm{c\overline{c}}}
\def\bbbar{\mathrm{b\overline{b}}}
\def\qqbar{\mathrm{q\overline{q}}}

\def\Zbb{\mathrm Z\arr \bbbar}
\def\bcl{\mathrm b \arr c \arr \ell}
\def\dmd{\Delta m_{\mathrm{d}}}
\def\dms{\Delta m_{\mathrm{s}}}
\def\dmq{\Delta m_{\mathrm{q}}}

\def\etal{{\sl et al.}}

\begin{document}           
\begin{titlepage}
\begin{center}
  {\large   EUROPEAN LABORATORY FOR PARTICLE PHYSICS }
\end{center}
\bigskip
\begin{tabbing}
\` CERN-PPE/97-064 \\
\` 5 June 1997 \\
\end{tabbing}
\vspace{10 mm}
\begin{center}{\LARGE\bf
An Updated Study of B Meson Oscillations using Dilepton Events
}\end{center}
\vspace{10 mm}
\begin{center}{\LARGE
The OPAL Collaboration
}\end{center}
\vspace{25mm}
\begin{abstract}
This paper reports a study of B meson oscillations
using hadronic $\Zzero$ decays with two identified leptons,
and updates a previous publication
by including data collected in 1994.
Decay times are reconstructed for each of the semileptonic B decays 
by forming
vertices which include the lepton 
and by estimating the B meson momentum.
The mass difference, $\dmd,$
between the two mass eigenstates in the $\Bd$ system
is measured to be 
$0.430 \pm 0.043 ~^{+0.028}_{-0.030}$~ps$^{-1}$,
where the first error is statistical and the second error
is systematic.
For the $\Bs$ system,
a lower limit of
$\dms > 2.2$~ps$^{-1}$
is obtained at 95\% C.L.
\vspace{10 mm}
\parskip 3cm
\end{abstract}
\vspace{35 mm} 
\begin{center}
(To be submitted to Zeitschrift f\"{u}r Physik C)
\end{center}
\end{titlepage}
%
%
\begin{center}{\Large        The OPAL Collaboration
}\end{center}\bigskip
\begin{center}{
K.\thinspace Ackerstaff$^{  8}$,
G.\thinspace Alexander$^{ 23}$,
J.\thinspace Allison$^{ 16}$,
N.\thinspace Altekamp$^{  5}$,
K.J.\thinspace Anderson$^{  9}$,
S.\thinspace Anderson$^{ 12}$,
S.\thinspace Arcelli$^{  2}$,
S.\thinspace Asai$^{ 24}$,
D.\thinspace Axen$^{ 29}$,
G.\thinspace Azuelos$^{ 18,  a}$
% ...
}\end{center}\bigskip
\bigskip
$^{  1}$School of Physics and Space Research, University of Birmingham,
Birmingham B15 2TT, UK
\newline
$^{  2}$Dipartimento di Fisica dell' Universit\`a di Bologna and INFN,
I-40126 Bologna, Italy
\newline
% ...
$^{  5}$Cavendish Laboratory, Cambridge CB3 0HE, UK
\newline
% ...
$^{  8}$CERN, European Organisation for Particle Physics,
CH-1211 Geneva 23, Switzerland
\newline
$^{  9}$Enrico Fermi Institute and Department of Physics,
University of Chicago, Chicago IL 60637, USA
\newline
% ...
$^{ 12}$Indiana University, Department of Physics,
Swain Hall West 117, Bloomington IN 47405, USA
\newline
% ...
$^{ 16}$Department of Physics, Schuster Laboratory, The University,
Manchester M13 9PL, UK
\newline
% ...
$^{ 18}$Laboratoire de Physique Nucl\'eaire, Universit\'e de Montr\'eal,
Montr\'eal, Quebec H3C 3J7, Canada
\newline
% ...
$^{ 23}$Department of Physics and Astronomy, Tel Aviv University,
Tel Aviv 69978, Israel
\newline
$^{ 24}$International Centre for Elementary Particle Physics and
Department of Physics, University of Tokyo, Tokyo 113, and
Kobe University, Kobe 657, Japan
\newline
% ...
$^{ 29}$University of British Columbia, Department of Physics,
Vancouver BC V6T 1Z1, Canada
\newline
% ...
$^{  a}$ and at TRIUMF, Vancouver, Canada V6T 2A3
\newline
% ...
\newline
%
\newpage
\section{Introduction}
In the Standard Model, a second-order weak transition transforms
neutral B mesons into their antiparticles~\cite{mixall}.  
The neutral B mesons
therefore oscillate between particle and antiparticle states before
decaying.  The frequency of the oscillation depends on
the top quark mass, the Cabibbo-Kobayashi-Maskawa
matrix elements, and meson decay constants.
By analogy with the $\mathrm{K^0}$ case and neglecting CP violation, 
the mass eigenstates, $|\mathrm{B_1} \rangle$ and $|\mathrm{B_2} \rangle$,
of $\Bq$ (q=d or s) can be described as follows:
\begin{eqnarray*}
|\mathrm{B_1} \rangle & = & \frac{1}{\sqrt{2}}
(|\Bq \rangle + |\Bqb \rangle ), \\
|\mathrm{B_2} \rangle & = & \frac{1}{\sqrt{2}}
(|\Bq \rangle - |\Bqb \rangle ).
\end{eqnarray*}
If a $\Bq$ is produced at time $t=0$, the
probabilities of having a $\Bq$ or a $\Bqb$ at proper
time $t$ are\footnote{The contribution of $\Delta\Gamma$, the difference
between the total decay widths of the mass eigenstates, to the
oscillations is expected to be negligible and has been ignored.}
\begin{eqnarray*}
P_{\Bq}(t) & = & \frac{1}{\tau} e^{-t/\tau} ~
\cos^2 \left(\frac{\dmq\, t}{2}\right) \\
P_{\Bqb}(t) & = & \frac{1}{\tau} e^{-t/\tau} ~
\sin^2 \left(\frac{\dmq\, t}{2}\right) 
\end{eqnarray*}
where $\tau$ is the $\Bq$ lifetime.
The frequency of the oscillation is given by $\dmq$,
the mass difference of the two mass eigenstates
($\dmq = m_{\mathrm{B}_1} - m_{\mathrm{B}_2}$).
For $\BdBd$ mixing, time-integrated
measurements from ARGUS and CLEO give 
$x_{\mathrm{d}}= \dmd\, \tau =0.67\pm 0.08$~\cite{ACmix,pdg}.  
Published measurements of the
frequency of $\BdBd$ oscillations made at LEP
are available using several different
techniques~\cite{LEPMIX,OPMIX12,OPMIX3,OPMIXL,ALEPHll}.
Lower limits on $\dms$ have been reported by
the ALEPH~\cite{ALEPHll,ALEPH2} and the 
OPAL~\cite{OPMIX3,OPMIXL} collaborations.
 
Extracting information on CKM matrix elements from the 
measurements of $\dmd$ and $\dms$ is prone to large uncertainties
due to poorly known meson decay constants.
These uncertainties can be reduced by considering the ratio
$\dms / \dmd$.
Given the present knowledge of $V_{\mathrm{ts}}$ and $V_{\mathrm{td}}$
one expects $\dms$ to be of the order of 
$10 \mathrm{~ps}^{-1}$~\cite{CKMconstraint}.
Using dilepton events in data collected between 
1991 and 1993~\cite{OPMIX3},
we studied $\BdBd$ and $\BsBs$ oscillations, 
reporting 
$\dmd = 0.462 ~^{+0.040}_{-0.053} ~^{+0.052}_{-0.035}$~ps$^{-1}$
and $\dms > 2.2$~ps$^{-1}$ at 95\% C.L.
%
We update these results by 
including data collected in 1994.
The technique is the same as that reported previously~\cite{OPMIX3}.
Hadronic $\Zzero$ decays with two lepton candidates, one in each
thrust hemisphere, are selected.
The reconstruction of a secondary vertex that includes the lepton is
attempted for each lepton candidate, yielding an estimate of the
decay length of the b hadron.
This is combined with an estimate of the relativistic boost of the
b hadron to give the proper decay time.
The likelihood of each event is calculated 
as a function of $\dmd$ and $\dms$
according to the 
measured proper times and the charge combination of the 
two leptons.
The result for $\dmd$ and the lower limit
on $\dms$ are then obtained
using a maximum likelihood technique.

\section{Event Selection and Simulation}

\subsection{Event Selection}
 

The analysis is performed on data collected by OPAL
in the vicinity of the
$\Zzero$ peak from 1991 to 1994. 
The OPAL detector has been described elsewhere~\cite{OPAL,opalsi}. 
Hadronic $\Zzero$ decays are selected using criteria
described in \cite{TKMH}.
A cone jet algorithm~\cite{conejet} is used to classify
tracks and electromagnetic clusters 
not associated to tracks into jets.
The size of the cone is chosen so as to include nearly all the decay
products of a b hadron into one jet.
The jets also include particles produced in the fragmentation process,
which originate from the 
$\mathrm{e^+e^-}$ collision point.
A total of
$2\,874\,660$ hadronic events satisfy the event selection criteria.
 
Electrons are identified using an artificial neural 
network~\cite{OPMIX3}
which is trained on a sample of 
simulated hadronic $\Zzero$ decays.
Electrons from photon conversions are rejected as in~\cite{Zpaper}.
Muons are identified as in~\cite{muextra}.
Lepton candidates are required to satisfy 
$p > 2.0\,$GeV
and \mbox{$|\cos \theta | < 0.9$}. 
Additional kinematic criteria are imposed
to reduce the fraction of
leptons in the sample coming from cascade decays of the type
$\mathrm{b\ra c\ra\ell}$.  
  
The techniques for secondary vertex reconstruction
and proper time estimation are described in \cite{OPMIX3}.
Dilepton
events with at least one reconstructed vertex are selected.

\subsection{Event Simulation}   
 
Monte Carlo events are used to predict the relative abundances and
decay time distributions for lepton candidates from various
physics processes.
The JETSET~7.4 Monte Carlo program~\cite{jetset}
with parameters tuned to OPAL data~\cite{jset2}
is used to generate $\Zzero\ra\qqbar$ events
which are processed by the detector simulation program~\cite{gopal}.
The fragmentation of b and c quarks is parametrised using
the fragmentation function of Peterson \etal~\cite{peterson},
with $\langle x_E\rangle$ 
for b and c hadrons given by the central values in Table~\ref{tab:mcpar}.
\begin{table}[htbp]
\begin{center}
\begin{tabular}{|c|c|} \hline
Quantity & Value \\ \hline
$\langle x_E\rangle_b$ &$0.697\pm 0.013$ \cite{muextra}\\
$\langle x_E\rangle_c$ &$0.51\pm 0.02$ \cite{Zpaper}\\
$B (\mathrm{ b\ra\ell} )$ &$(10.5\pm 0.6 \pm 0.5)\%$ \cite{muextra}\\
$B (\mathrm{ b\ra c\ra\ell} )$ &$(7.7\pm 0.4 \pm 0.7)\%$ \cite{muextra}\\
$B (\mathrm{ b\ra \bar{c}\ra\ell} )$ 
 &$(1.3\pm 0.5)\%$ \cite{muextra}\\
$M \mathrm(\Bs )$ & 5.48 $\mathrm{GeV}$ \\
$M (\mathrm{\Lambda_b})$ & 5.62 $\mathrm{GeV}$ \\
$\tau_{\BPLUS}/\tau_{\Bd}$ & $1.03\pm 0.06$ \cite{pdg}\\
$\tau_{\Bs}/\tau_{\Bd}$ & $1.03\pm 0.08$ \cite{pdg}\\
$\tau_{\mathrm{\Lambda_b}}/\tau_{\Bd}$
 & $0.73\pm 0.06$ \cite{pdg}\\
$\mathrm{ \langle \tau_b \rangle}$ &$1.55 \pm 0.02$ ps
~\cite{pdg}\\ \hline
\end{tabular}
 
\caption{The parameters used for the Monte Carlo simulation.}
\label{tab:mcpar}
\end{center}
\end{table}
 
Standard Model values of the partial widths of the $\Zzero$
into $\qqbar$ are used~\cite{Stan}.
The mixture of c-flavoured hadrons produced both in $\Zzero\ra\ccbar$
events and in b hadron decays is as 
prescribed in \cite{Zpaper}.
The semileptonic branching ratios of charm hadrons and associated
uncertainties are also those 
of \cite{Zpaper}.
The central values in Table~\ref{tab:mcpar} are taken for the
inclusive branching ratios for $\mathrm{b\ra\ell}$,
$\mathrm{b\ra c\ra\ell}$ and $\mathrm{b \ra \bar{c} \ra \ell}$. 
The semileptonic branching ratios
of the individual b hadrons are assumed to be proportional to
the lifetimes.
The models used in describing the semileptonic decays of b and c hadrons
are those used in determining the central values in \cite{Zpaper}.
The asssumed masses for $\Bs$ and $\mathrm{\Lambda_{b}}$ particles 
are also given in Table~\ref{tab:mcpar}.
The lifetimes of b hadrons used in this analysis 
were taken from the world average values~\cite{pdg}, as
indicated in Table~\ref{tab:mcpar}.
 
%
\section{Fit Results for \mbox{\boldmath $\dmd$}}

The numbers of dilepton events with at least one secondary vertex
constructed for the combination of e-e, e-$\mu$ and $\mu$-$\mu$,
are listed in Table~\ref{tab:number},
separately for like-sign and unlike-sign dilepton events.
Also included is the total number of secondary vertices reconstructed 
in these events.
\begin{table}[htbp]
\begin{center}
\begin{tabular}{|l||c|c|c||c|c|} \hline
            &  e-e & e-$\mu$ & $\mu$-$\mu$ &
               total & total vertices \\\hline\hline
unlike-sign &  891  &  1791  &  1070  &  3752  &  5971 \\\hline
like-sign   &  377  &   780  &   448  &  1605  &  2573 \\\hline
\end{tabular}
\caption{
The numbers of dilepton events with at least one secondary vertex
reconstructed for the combinations
e-e, e-$\mu$ and $\mu$-$\mu$, separately
for like-sign and unlike-sign leptons.
Also indicated is the total number of secondary vertices reconstructed
in unlike-sign and like-sign dilepton events.
}
\label{tab:number}
\end{center}
\end{table}
 
\begin{figure}[htbp]
\centering
\epsfxsize=16cm
\begin{center}
    \leavevmode
    \epsffile[14 137 581 754]{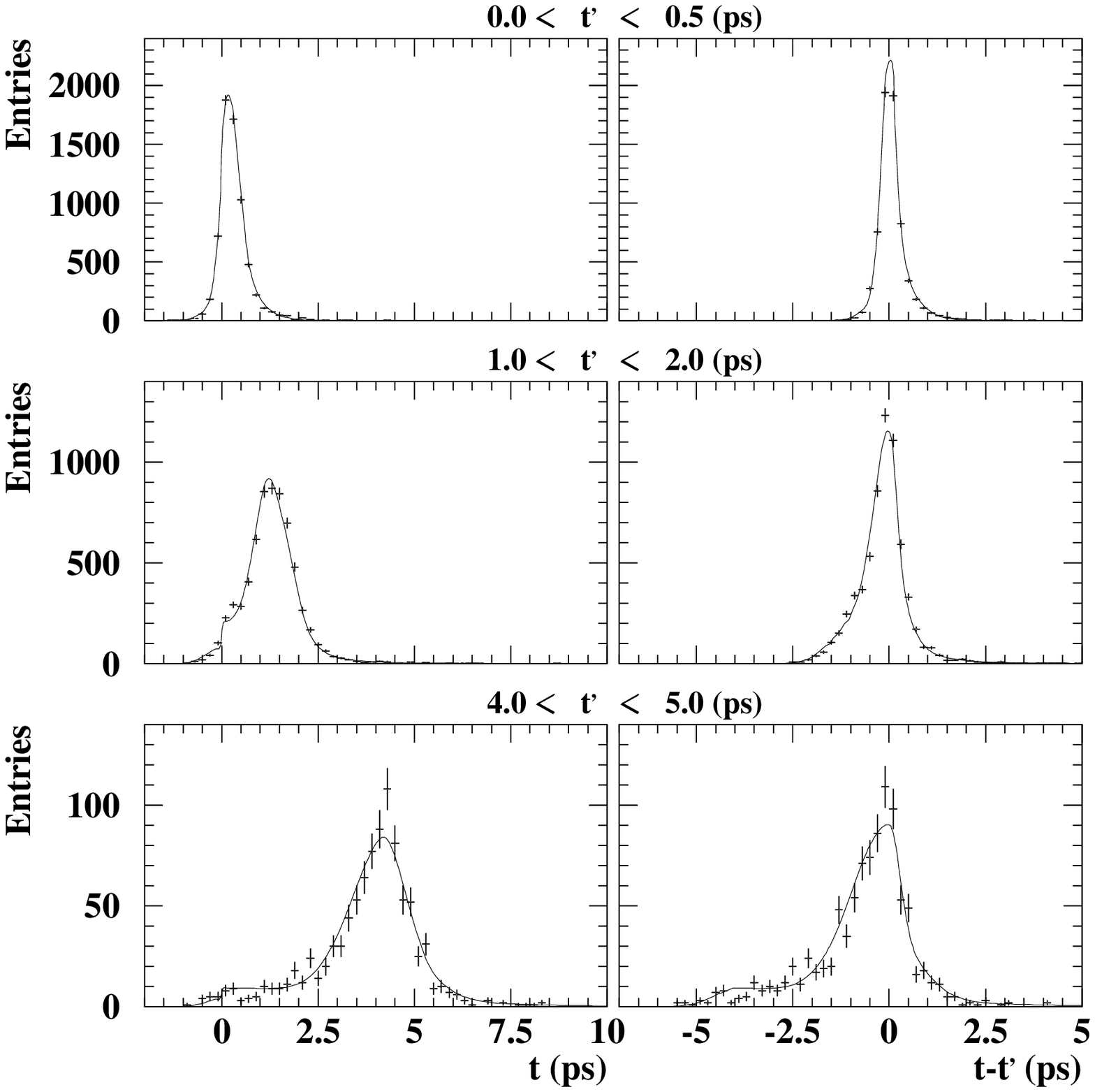}
\end{center}
\caption{
The distributions of reconstructed proper time, $t$, and $t-t'$
in three slices of the true proper time $t'$ for 
leptons from primary b hadron decays in the Monte Carlo.
Also shown is the parametrisation of these distributions.
}
\label{fig:proj}
\end{figure}

In order to study $\dmd$ and $\dms$, the likelihood
of the event sample is calculated as a function of these parameters.
%
The construction of the likelihood function 
follows the procedure described in
the previous paper~\cite{OPMIX3}.
%
The true proper time distribution 
is described by a physics function
for each source of events.
The B mixing is also described by the physics function.
The reconstructed time distributions, $f(t)$, are then obtained by
convolving the physics function with resolution functions, $P(t,t')$,
which describe the proper time resolution for each source.
For example, in the absence of mixing,
\[
f(t) = \frac{1}{\tau}\int_{0}^{\infty} e^{-\frac{t'}{\tau}} P(t,t')
{\mathrm d}t'
\]
for a source with lifetime $\tau$.
The resolution functions are complicated functions, which
must include a description of misreconstruction near $t=0$,
even when the true proper time is large.
The probability of this misreconstruction depends on
the true proper time,
and the description of the resolution function was modified
from the previous paper to describe this better.
%
The resolution function has the following form:
\begin{eqnarray*}
P(t,t') = C[(1-\exp{(-\frac{t'}{\alpha})}) v(t,t') +
\exp{(-\frac{t'}{\alpha})} u(t)]
\end{eqnarray*}
where $t$ and $t'$ are the reconstructed proper time and
the true proper time, respectively.
The functions
$v(t,t')$ and $u(t)$ 
describe the reconstructed proper time distributions for
the correctly reconstructed and misreconstructed vertices
respectively.
The parameter $C$ is a normalization factor, while
$\alpha$ is a parameter to describe the 
dependence of the misreconstruction probability
on the true proper time.
Distributions of 
$t$ and $t-t'$ are shown
in Figure~\ref{fig:proj}
for three slices of true proper time $t'$ for Monte Carlo
b decays.
The fitted resolution function, which is superimposed
in the figure, 
describes the distributions well.
%
 
To determine $\dmd$ a three parameter fit is performed, varying $\dmd$ 
simultaneously with 
the cascade fraction,
the fraction of lepton candidates in $\Zbb$ decays
that are due to $\bcl$ decays,
and the $\Bs$ production fraction, the fraction
of b quarks that give rise to $\Bs$ mesons. 
Gaussian constraints reflecting the systematic errors 
on these two parameters are imposed. 
The relative uncertainty in the cascade fraction
is taken to be $\pm 15\%$~\cite{Zpaper}, which includes
uncertainties due to
branching fractions, decay modelling and detector simulation.
%
The $\Bs$ production fraction, $f_{\mathrm s}$, is constrained both by direct
measurements, giving a rate of $(11.1\pm 2.6)$\% \cite{pdg} relative to
all weakly decaying b hadrons, and by the measured average mixing rate
of b hadrons, $\bar{\chi} = 0.126\pm 0.008$ \cite{pdg}
together with knowledge
of the equivalent parameters, $\chi_{\mathrm d}$ and $\chi_{\mathrm s}$, 
for $\Bd$ and $\Bs$ mesons
($\chi _{\mathrm d(s)} = 0.5 \times 
 x_{\mathrm d(s)}^2/(1+x_{\mathrm d(s)}^2)$).
This is equivalent to the constraint 
$f_{\mathrm s} = (11.2^{+1.8}_{-1.9})$\% \cite{pdg}
except that the values of $\chi_{\mathrm d}$ and $\chi_{\mathrm s}$
are calculated from the values of $\dmd$ and $\dms$ in the fit,
together
with the appropriate lifetimes.

In the fit the $\Bs$ oscillation parameter is fixed at  
$\dms = 10.0$~ps$^{-1}$.
The result of the fit is 
$\dmd = 0.430 \pm 0.045$~ps$^{-1}$.     
The fitted value of the cascade fraction is 
$0.078 \pm 0.007$
compared to the nominal value of 0.066,
as calculated from the Monte Carlo sample.
The fitted value of $f_{\mathrm s}$ 
is $(12.7 \pm 1.9)$\%.
  
\begin{figure}[htbp]
\centering
\epsfxsize=16cm
\begin{center}
    \leavevmode
    \epsffile[14 137 581 704]{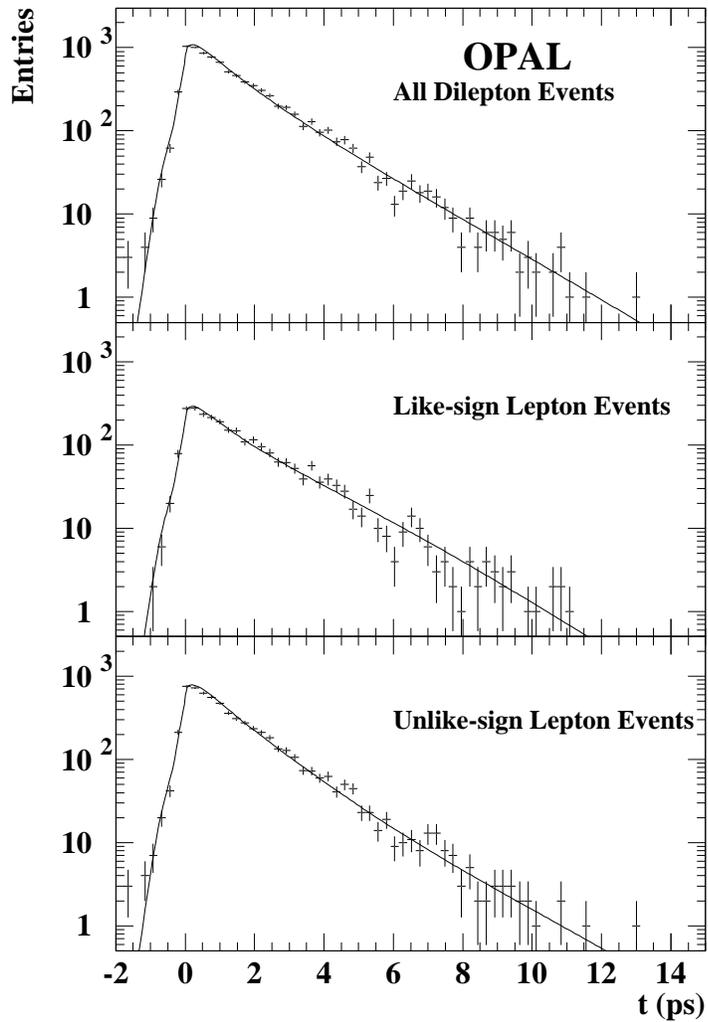}
\end{center}
\caption{
The proper time distributions for all leptons in dilepton
events (top) for which a vertex is
found, and for those leptons in like-sign (centre) and
unlike-sign (bottom) events.  The
curves represent the results of the maximum likelihood fit.
}
\label{fig:distriobs}
\end{figure}

Figure~\ref{fig:distriobs} shows the distribution
of decay times for all leptons in the
dilepton sample and separately for leptons in like-sign and unlike-sign
events.
The curves are the results of the likelihood fit.
  
\begin{figure}[htbp]
\centering
\epsfxsize=16cm
\begin{center}
    \leavevmode
    \epsffile[14 137 581 704]{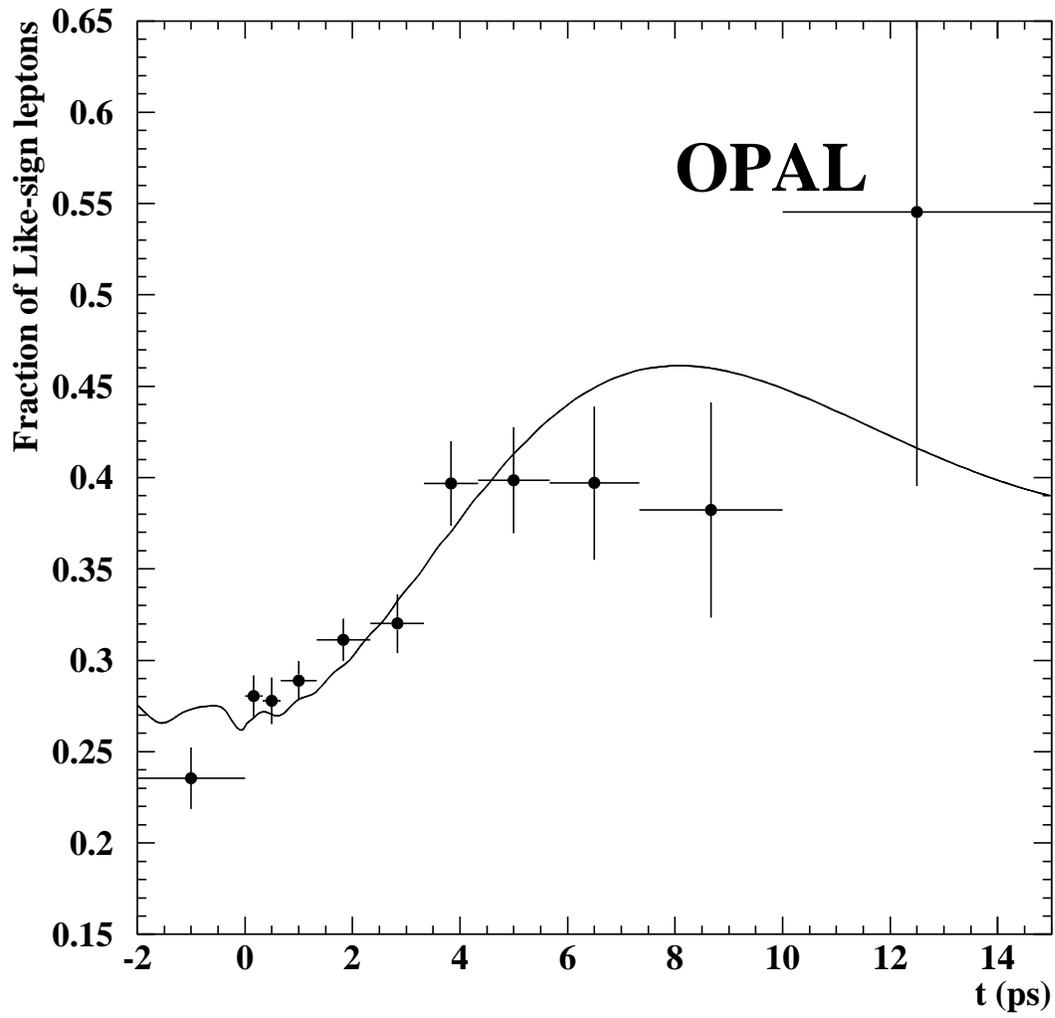}
\end{center}
\caption{
The fraction of like-sign leptons as a function of proper decay time:
$\cal{R}$$(t)$.
The solid curve represents the expectation with $\dmd$ set to
0.430~ps$^{-1}$ and $\dms$ set to 10.0~ps$^{-1}$.
}
\label{fig:folevent}
\end{figure}

The fraction of like-sign leptons as a function of proper decay time,
\begin{eqnarray*}
{\cal{R}}(t) = \frac{ N^{LS} (t)  }
{ N^{US} (t) + N^{LS} (t) } \; ,
\end{eqnarray*}
is plotted in Figure~\ref{fig:folevent} for data,
where $N^{LS} (t)$ ($N^{US} (t)$) is the reconstructed time
distribution for leptons in like-sign (unlike-sign) events.
In the figure, the expected curve for $\dmd=0.430$~ps$^{-1}$ is shown
as the solid line.
The fitted values of the cascade fraction 
and the fraction of leptons from $\Bs$ 
decays are used.
Events in which vertices have
been reconstructed in both thrust hemispheres 
enter the plot twice.

%
\section{Systematic Errors on \mbox{\boldmath $\dmd$}}

In the three parameter fit, the error on $\dmd$ is a
combination of statistical error and
systematic error due to the constraints on
the cascade decay fraction and the $\Bs$ production fraction.
The systematic error from the cascade decay fraction is estimated
by repeating the fit with
the central value of the constraint
for the cascade fraction changed by $+15$\% or
$-15$\% (the systematic uncertainty on this parameter) from
its nominal value.
The systematic error resulting
from the $\Bs$ production fraction is obtained in a similar
way.
The statistical error on $\dmd$ is
$\pm 0.043$~ps$^{-1}$, obtained by subtracting in quadrature
these two systematic errors from the fit error.
 
The uncertainty due to the resolution function description is assessed 
by repeating the parametrisation using Monte Carlo events 
in which the tracking 
resolution is degraded by 10\% \cite{gbb} or improved by 10\%. 
The uncertainty in the background from $\Zzero\ra\ccbar$ events is taken
to be $\pm 30\%$ due to uncertainties in the branching fractions and
modelling of semileptonic charm decays, the relative production rates
of charmed hadrons, and the uncertainty in the partial width for
$\Zzero\ra\ccbar$.    
The production rates of $\Bd$ and $\BPLUS$ are assumed to be equal
and the b-baryon production rate
is assumed to lie in the range $(9 \pm 4)$\%. 
The fraction of $\mathrm D^{**}$ produced in decays of b hadrons was
assumed to be
$B(\mathrm b \arr D^{**}) = 0.36 \pm 0.10$.
Variations in the efficiency of the secondary vertex reconstruction
as a function of decay length are found to have a negligible
effect on $\dmd$. 
Uncertainties in the source composition due to Monte Carlo statistics
are negligible.
The B lifetime variations are performed by changing the ratios
of individual B lifetimes while keeping the average lifetime
fixed at the LEP, CDF and SLD average value, 1.55~ps$^{-1}$.
 
The summary of the sources and estimated values of
systematic errors
is given in Table~\ref{tab:Bdsys}.
The sum of these systematic errors in quadrature is
$\delta \dmd = ^{+0.028}_{-0.030}$~ps$^{-1}$.
 
\begin{table}[htbp]
\begin{center}
\begin{tabular}{|c|c|} \hline
Source of uncertainty and range & $\delta\dmd$~ps$^{-1}$ \\\hline\hline
cascade decay fraction ($\pm$15\%)  & $-0.010$ $+0.011$  \\ \hline
$\Bs$ fraction (see text)           & $-0.006$ $+0.006$  \\ \hline
resolution function ($\pm 10\%$)    & $+0.010$ $-0.010$  \\ \hline
lepton misidentification (e$:\pm 30\%, \mu:\pm 20\%$)
                                    & $-0.001$ $+0.000$  \\ \hline
charm background ($\pm 30\%$)       & $-0.006$ $+0.001$   \\ \hline
b-baryon fraction  ($\pm$0.04)      & $+0.010$ $-0.009$  \\ \hline
$B(\mathrm b \arr D^{**})$ ($\pm 0.10$)   & $+0.003$ $-0.003$ \\ \hline
$\tau_{\BPLUS} / \tau_{\Bd} = 1.03\pm 0.06$ & $+0.019$ $-0.023$ \\
$\tau_{\Bs} / \tau_{\Bd} = 1.03 \pm 0.08$
                               & $-0.000$ $+0.001$ \\
$\tau_{\mathrm{\Lambda_b}} / \tau_{\Bd} = 0.73\pm 0.06$
                               & $+0.004$ $-0.004$ \\ \hline
$\dms = 2-20$~ps$^{-1}$        & $+0.004$ $-0.000$ \\ \hline\hline
Total systematic error         & $+0.028$ $-0.030$ \\ \hline
\end{tabular}
\caption{
Summary of systematic errors on the $\dmd$
measurement.
For each source of error, the first error quoted results
from varying the parameter in the positive sense. 
}
\label{tab:Bdsys}
\end{center}
\end{table}

\section{Fit Results for \mbox{\boldmath $\dms$}} \label{dmsfit}

\begin{figure}[htbp]
\centering
\epsfxsize=16cm
\begin{center}
    \leavevmode
    \epsffile[14 137 581 704]{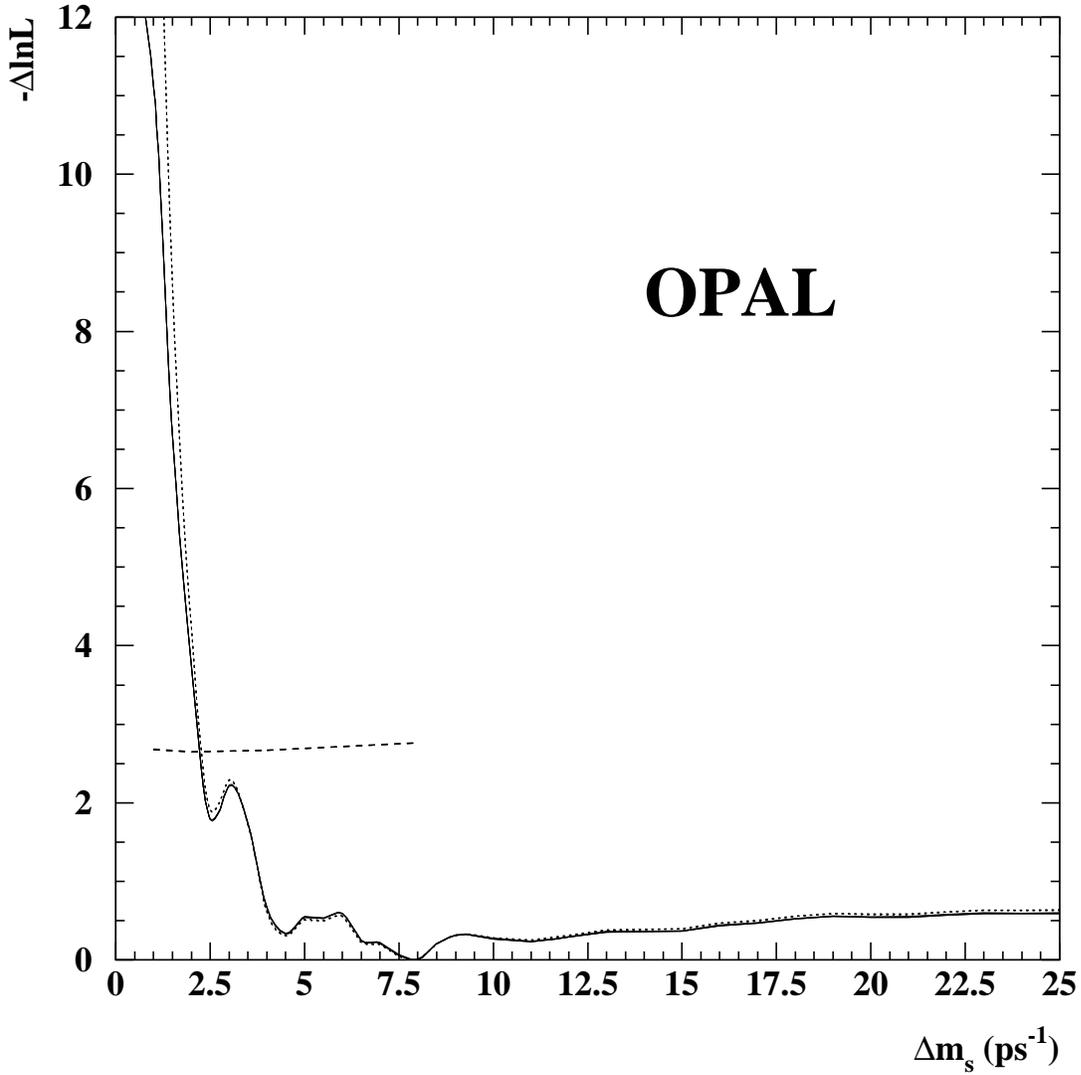}
\end{center}
\caption{
The difference in log-likelihood from the maximum value is
shown as a function of $\dms$.  The solid curve includes the effect of
systematic errors, while the dotted curve includes only
statistical errors. The dashed curve shows the 95\% C.L.
}
\label{fig:Bslimit}
\end{figure}

We use the dependence of the likelihood on the assumed value
of $\dms$
to derive a lower limit.
%
To account for systematic errors when setting the limit,
we produce a likelihood curve as a function of $\dms$ that
includes these effects.
This is achieved by maximizing the likelihood
with respect to the value of each relevant parameter,
constrained by a Gaussian error corresponding to its uncertainty,
at each value of $\dms$.
The constraints are those shown in Table~\ref{tab:Bdsys}.
In addition, $\dmd$ is treated as a systematic uncertainty
constrained by the average $\dmd$ from analyses using 
reconstructed D$^*$ mesons:
$\dmd = 0.52\pm0.05$~ps$^{-1}$~\cite{LEPMIX,OPMIX12}.
The exception to this scheme is the treatment of the resolution function
description.
In this case, three curves of $\ln\cal{L}$ were calculated, including
all other systematic uncertainties, assuming the default tracking resolution
or assuming the tracking resolution was degraded or improved by 10\%.
The smallest of the three values of $-\Delta\ln\cal{L}$ was taken
at each value of $\dms$.
The solid curve in
Figure~\ref{fig:Bslimit} shows the difference in
log-likelihood
from the maximum as a function of $\dms$ with
systematic errors included.
 
We set the limit on the basis of the
difference in log-likelihood, $\Delta\ln\cal{L}$, 
with respect to the maximum value.
A Monte Carlo technique was used to determine the correspondence
between confidence levels and values of $\Delta\ln\cal{L}$
as a function of $\dms$.
This approach is found to be more reliable than the approach of 
the previous paper~\cite{OPMIX3}, where $-\Delta\ln{\cal{L}} = 1.92$
was assumed to correspond to 95\% confidence level.
Many data sets, of the same size as the real data sample, were simulated
using a fast Monte Carlo and fitted in a manner similar to the data.
The main systematic errors affecting the $\dms$ result were simulated
by allowing the parameters of the Monte Carlo to vary independently
for each data set.
The corresponding parameters were allowed to vary under Gaussian constraints
in the fit.
The exception to this was the parameter governing the proper time resolution,
which was treated in the same way as in the data fit.
For each simulated
data sample a single value of 
$\Delta\ln{\cal{L}} = \ln{\cal{L}}_{\mathrm max} - \ln{\cal{L}}(\dms^*)$ 
was extracted, where $\dms^*$
is the generated value of $\dms$.
The $\Delta\ln\cal{L}$ corresponding to 95\% confidence was defined
to be that value above which lay only 5\% of the simulated data samples.
This procedure was performed for input values 
of $\dms^*=1.0, 2.0, 4.0$ and 8.0 ps$^{-1}$, using 3000 Monte Carlo
data sets at each value of $\dms^*$.
The results of this study are shown as the dashed line in 
Figure~\ref{fig:Bslimit}.
We exclude the region of $\dms < 2.2$~ps$^{-1}$ at 95\% C.L.

To assess the importance of the systematic errors, 
we studied the log-likelihood
as a function of $\dms$, while fixing all other parameters.
These parameters were set 
to the values that maximized 
the log-likelihood at the preferred value of 
$\dms$ in the procedure described above.
The result is shown as 
the dotted curve in Figure~\ref{fig:Bslimit}.
We conclude that the systematic errors have only a minor effect on our
result.
 
\begin{figure}[htbp]
\centering
\epsfxsize=16cm
\begin{center}
    \leavevmode
    \epsffile[14 137 581 704]{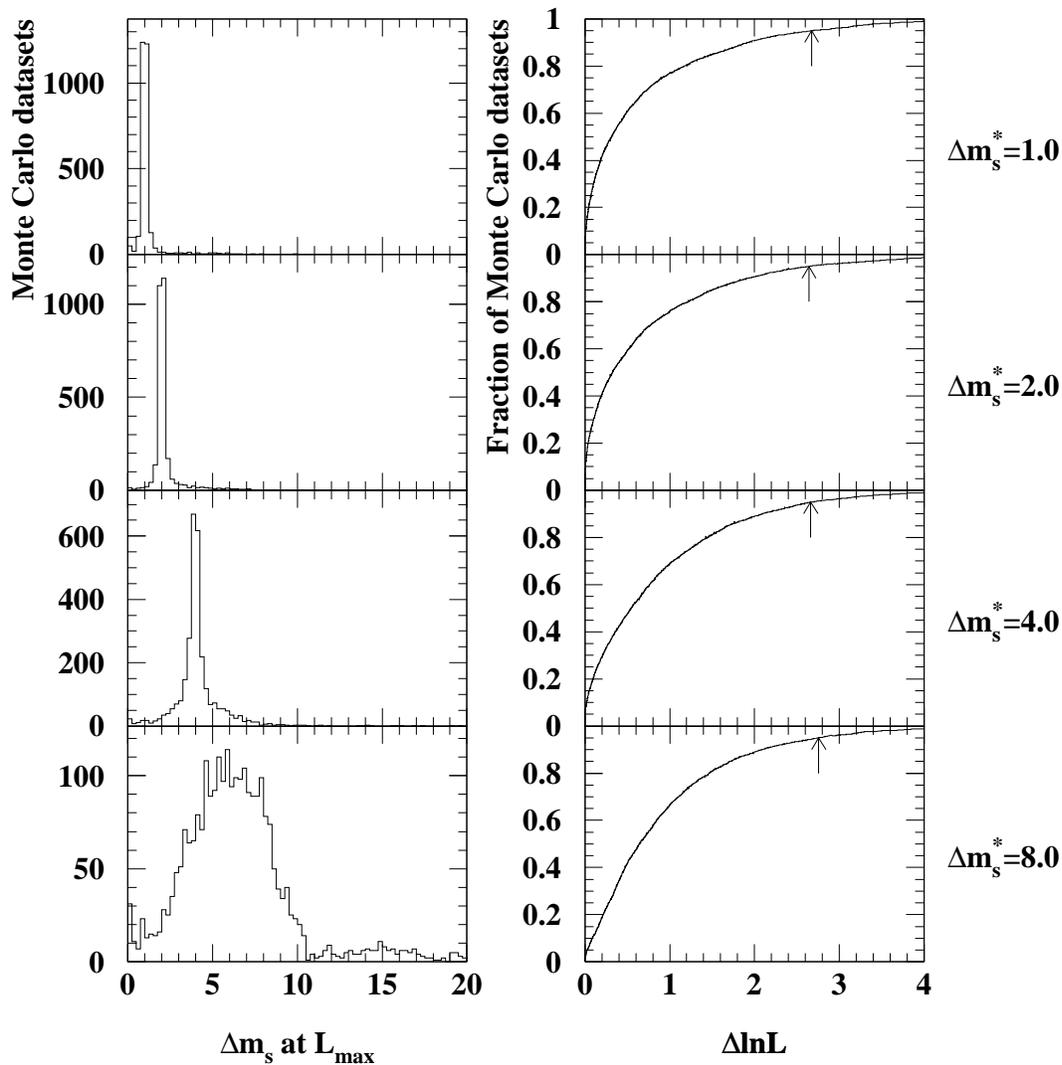}
\end{center}
\caption{
The results of fits to 3000 toy Monte Carlo datasets are shown.  The
$\dms^*$ value indicates the generated value of $\dms$ in ps$^{-1}$. 
The left-hand column shows the distribution of 
fitted values of $\dms$ for each $\dms^*$ value.
The right-hand column shows the normalised cumulative distribution
of the difference between the log-likelihood values at the
fitted maximum and at the generated value.
The arrow indicates the 95\% confidence level value.
}
\label{fig:TOYMCresult}
\end{figure}

Using the data sets simulated with the fast Monte Carlo referred to above,
we were able to check the analysis technique and study the
expected sensitivity to $\dms$.
The results of these studies are shown
in Figure~\ref{fig:TOYMCresult}.
Each row corresponds to a different generated value
of $\dms$, indicated by the $\dms^*$ at the right edge of the plot.
The left column shows the fitted value of $\dms$ for each trial.
In the right column are normalised cumulative
distributions of the log-likelihood
difference between the value at the fitted maximum and the value at
the generated $\dms$ for each trial.
The sensitivity of this analysis is good for $\dms^* < 4$~ps$^{-1}$,
but is lost
between 4~ps$^{-1}$ and 8~ps$^{-1}$.

\section{Conclusion}
We have measured the oscillation frequency $\dmd$
by measuring the proper time of B meson decays
and tagging the charges of
leptons in both thrust hemispheres.
The $\BdBd$ oscillation parameter 
is measured to be:
$$\dmd = 0.430 \pm 0.043 ~^{+0.028}_{-0.030}~\mathrm{ps}^{-1} \; ,$$
corresponding to 
$(2.83 \pm 0.28 ^{+ 0.18}_{-0.20})\times 10^{-4}$ eV.
This result is consistent with and supersedes the result using 1991-1993 data.
%
%
 

The $\dmd$ value is consistent with the OPAL results
$\dmd = 0.548 \pm 0.050 ~^{+0.023}_{-0.019}$~ps$^{-1}$~\cite{OPMIX12}
from data containing $\DSPM$ mesons and leptons,
and $\dmd = 0.444 \pm 0.029
~^{+0.020}_{-0.017}$~ps$^{-1}$~\cite{OPMIXL}
from inclusive lepton events.
Combining these results, taking into account
correlations in the systematic errors, we find
$$\dmd = 0.467\pm 0.022 ~^{+0.017}_{-0.015}~\mathrm{ps}^{-1} \; . $$
The small statistical correlations between the results were
found to have a negligible effect.
This result
is consistent with previous measurements~\cite{LEPMIX,ALEPHll}.
Using $\tau_{\Bd}=1.56\pm 0.06\,\mathrm{ps}$, the combined OPAL
value gives
$x_{\mathrm d} = 0.73\pm 0.04\pm 0.03$, where the last error
is due to the uncertainty in $\tau_{\Bd}$.  
This value is also consistent with the average of
ARGUS and CLEO measurements,
$x_{\mathrm d} = 0.67\pm 0.08$~\cite{ACmix,pdg}.

We obtain a lower limit on $\dms$ at 95\% confidence level:
$\dms > 2.2~\mathrm{ps}^{-1} .$
This limit is less constraining than the ALEPH
results\cite{ALEPHll,ALEPH2}
and a recent OPAL result~\cite{OPMIXL}.
%
\par
\vspace*{1.cm}
\section*{Acknowledgements}
\noindent
We particularly wish to thank the SL Division for the efficient operation
of the LEP accelerator and for their continuing close cooperation with
our experimental group. We thank our colleagues from CEA, DAPNIA/SPP,
CE-Saclay for their efforts over the years on the time-of-flight and trigger
systems which we continue to use.  In addition to the support staff at our own
institutions we are pleased to acknowledge the  \\
Department of Energy, USA, \\
National Science Foundation, USA, \\
Particle Physics and Astronomy Research Council, UK, \\
Natural Sciences and Engineering Research Council, Canada, \\
Israel Science Foundation, administered by the Israel
Academy of Science and Humanities, \\
Minerva Gesellschaft, \\
Benoziyo Center for High Energy Physics,\\
Japanese Ministry of Education, Science and Culture (the
Monbusho) and a grant under the Monbusho International
Science Research Program,\\
German Israeli Bi-national Science Foundation (GIF), \\
Direction des Sciences de la Mati\`ere du Commissariat \`a l'Energie
Atomique, France, \\
Bundesministerium f\"ur Bildung, Wissenschaft,
Forschung und Technologie, Germany, \\
National Research Council of Canada, \\
Hungarian Foundation for Scientific Research, OTKA T-016660,
T023793 and OTKA F-023259.\\

%
%
\newpage
%
%
\newpage
%
%
\newpage
%
%
\newpage
%
%
%
%
\newpage

\end{document}